\documentclass[twocolumn,twocolappendix,apj]{aastex63_davidehack}

\usepackage{amssymb,amsmath,verbatim,mathtools,needspace,enumitem,graphicx,physics,microtype,mathrsfs,amssymb,xcolor,afterpage}

\def\apjl{\rm{ApJ}}
\def\apjs{\rm{ApJS}}
\def\aap{\rm{A\&A}}

\newcommand{\ssim}{\mathchar"5218\relax\,}

\newcommand{\bham}{\affiliation{School of Physics and Astronomy \& Institute for Gravitational Wave Astronomy, \\ University of Birmingham, Birmingham, B15 2TT, United Kingdom}}

\newcommand\orcid[1]{\href{https://orcid.org/#1}{$\!\!$\includegraphics[scale=0.006]{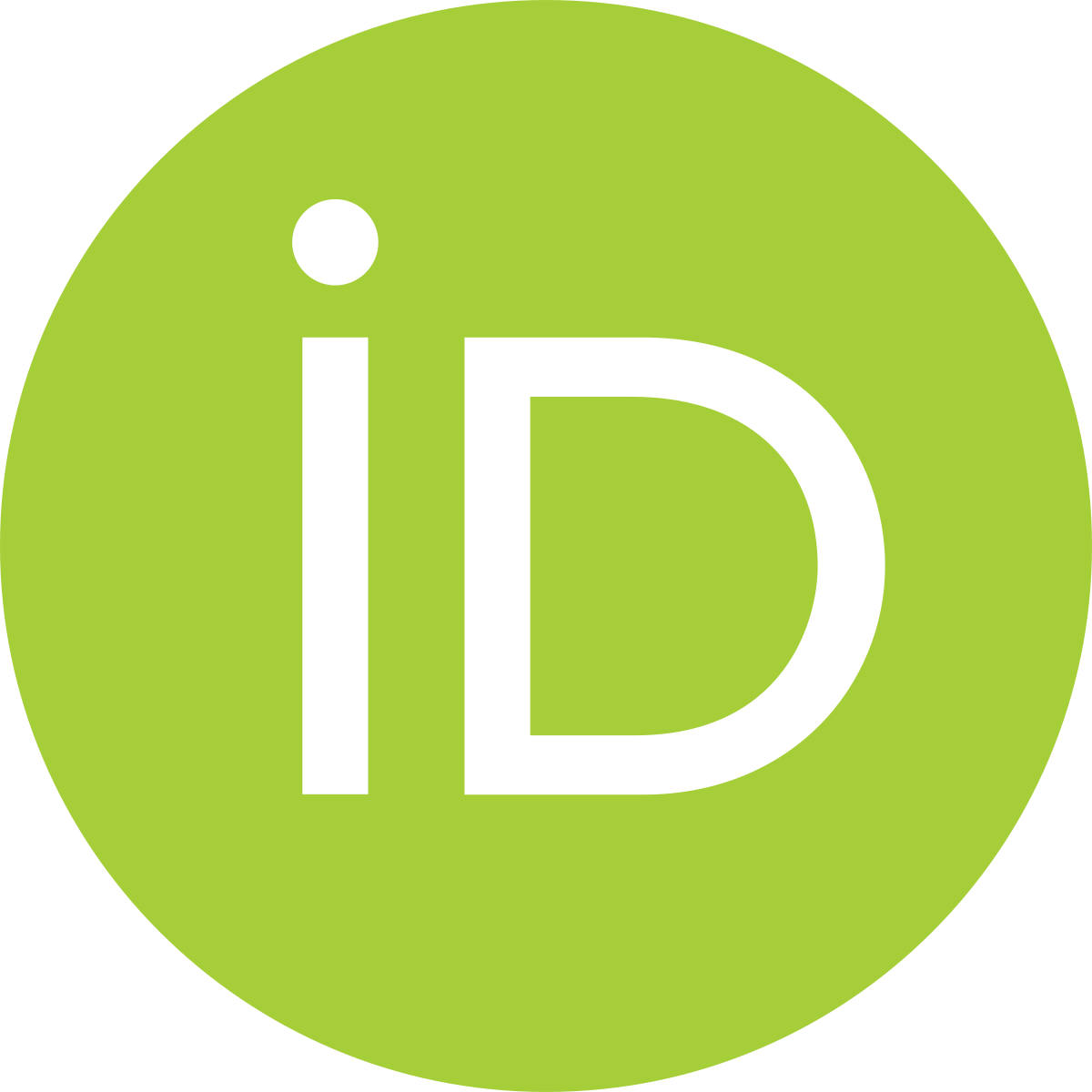} $\!\!$}}

\begin{document}

\title{High mass but low spin: an exclusion region to rule out hierarchical black-hole \\  mergers  as a mechanism to populate the pair-instability mass gap}

\author{Davide Gerosa \orcid{0000-0002-0933-3579}}
\email{d.gerosa@bham.ac.uk}
 \bham

\author{Nicola Giacobbo \orcid{0000-0002-8339-0889}}
 \bham

\author{Alberto Vecchio \orcid{0000-0002-6254-1617}}
 \bham
\date{\today}

\begin{abstract}
The occurrence of pair-instability supernovae is predicted to prevent the formation of black holes with masses $\gtrsim 50 M_\odot$. Recent gravitational-wave detections in this mass range require an explanation beyond that of standard stellar collapse. Current modeling strategies 
include the hierarchical assembly of previous generations of black-hole mergers as well as other mechanisms of astrophysical nature (lowered nuclear-reaction rates, envelope retention, stellar mergers, accretion, dredge-up episodes). %
In this paper, we point out the occurrence of an exclusion region that cannot be easily populated by hierarchical black-hole mergers. A future gravitational-wave detection of a black hole with mass $\gtrsim 50M_\odot$ and spin $\lesssim 0.2$ will indicate that the pair-instability mass gap is polluted in some other way. Such a putative outlier can be explained using hierarchical mergers  only with considerable fine-tuning of both mass ratio and spins of the preceding black-hole merger ---an assumption that can then be cross-checked against the bulk of the gravitational-wave catalog.  
\end{abstract}


\section{Introduction}

The black-hole (BH) pair-instability mass gap is one of the cleanest predictions that stellar astrophysics has provided to the nascent field of gravitational-wave (GW) astronomy. If the helium core of a collapsing star becomes unstable to pair production, the lack of the necessary pressure ignites carbon and oxygen in an explosive fashion, disrupting the remnant~\citep{1964ApJS....9..201F,1967PhRvL..18..379B}. %
The key prediction is that BHs in a precise mass range should be substantially rarer, or even forbidden~\citep{2016A&A...594A..97B,2016ApJ...824L..10W,2019ApJ...882..121S}. Such a ``gap'' in the BH mass spectrum extends from $\ssim 50 M_\odot$ to $\ssim 120 M_\odot$ \citep{2019ApJ...887...53F,2020MNRAS.493.4333R} and offers precious opportunities to test the underlying astrophysical processes with  GW data.

The occurrence of an upper mass cutoff in the population of merging BHs is now a solid observational finding. Current GW data indicate the BH-binary merger rate  decreases  drastically for masses $\gtrsim 40 M_\odot$~\citep{2021ApJ...913L...7A}. At the same time, some of the detected BHs are compatible with masses above this threshold. The most striking case is the primary component of GW190521, which has a mass $>65 M_\odot$ at $>99.5\%$ confidence~\citep{2020PhRvL.125j1102A}. GW observations are telling us that the pair-instability mass gap exists, but some BHs can ``fill'' it.

The hierarchical assembly of multiple generations of BH mergers is a compelling explanation for such heavy outliers~(\citealt{2017PhRvD..95l4046G,2017ApJ...840L..24F}; for a review see \citealt{2021arXiv210503439G}). In this scenario, the observed BHs do not originate directly from the collapse of stars, but rather from the merger of previous generations of BH binaries. Several astrophysical environments have been put forward as suitable hosts of hierarchical mergers  \citep {2016ApJ...824L..12O,2019PhRvD.100d3027R,2019PhRvL.123r1101Y,2020PhRvD.102d3002B,2020ApJ...898...25T,2021arXiv210305016M}. Among those, nuclear star clusters (e.g.
\citealt{2009ApJ...692..917M,2009MNRAS.395.2127O}) and the disks of active galactic nuclei (AGN; e.g.~\citealt{2017ApJ...835..165B,2017MNRAS.464..946S})  appear to be particularly promising because their escape speeds exceed that of typical recoils BHs receive at merger~\citep{2019PhRvD.100d1301G}. %

There are, however, other astrophysical mechanisms one can invoke to produce BHs heavier than $\ssim 50 M_\odot$. The mass of a stellar-origin BH strongly depends on the details of the evolution of its progenitor star and many evolutionary processes are still uncertain. Recent studies have shown that one can populate the mass gap by exploiting uncertainties in the $^{12}$C($\alpha, \gamma $)$^{16}$O reaction rate  \citep{2018ApJ...863..153T,2020ApJ...905L..15B,2020ApJ...902L..36F,2021MNRAS.501.4514C,2021ApJ...912L..31W}, as well as by reviewing stellar-wind prescriptions~\citep{2019ApJ...887...72L,2020ApJ...890..113B,2021MNRAS.504..146V} and including stellar rotation \citep{2020A&A...640L..18M,2021ApJ...912L..31W}. The presence of stellar companions and/or gaseous environments might also aid the formation of heavier BHs via either stellar mergers \citep{2019MNRAS.487.2947D,2020MNRAS.497.1043D,2019MNRAS.485..889S,2020ApJ...903...45K,2020ApJ...904L..13R,2021ApJ...908L..29G} or accretion episodes~\citep{2019ApJ...882...36M,2019A&A...632L...8R,2020ApJ...903L..21S,2020ApJ...897..100V,2021MNRAS.501.1413N,2021ApJ...908...59R}. Forming BHs from Population III also provides an appealing explanation to the larger masses involved~\citep{2020ApJ...903L..40L,2020ApJ...892...41S, 2020arXiv201007616T,2021ApJ...910...30T,2020ApJ...905L..21U,2021MNRAS.502L..40F,2021MNRAS.501L..49K}.
More speculative explanations include primordial BHs~\citep{2021PhRvL.126e1101D}, exotic compact objects ~\citep{2021PhRvL.126h1101B}, and physics beyond the standard model~\citep{2020PhRvL.125z1105S}.

In this paper, we explore the distinguishability of hierarchical BH mergers as a mechanism to populate the  pair-instability mass gap. In particular, we point out the existence of an exclusion region: objects with \emph{both large masses 
and small spins} cannot be easily produced by either conventional stellar collapse or hierarchical BH mergers. If a future LIGO/Virgo observing run delivers a BH with mass $m\gtrsim 50M_\odot$ but dimensionless spin $\chi\lesssim 0.2$, such an event will need to be explained with other processes.  This would imply that the astrophysics of BH-binary formation does not, after all, impose a strict limit to the mass of the remnant. On the other hand, observing several mass-gap events all with large spins would stress that hierarchical mergers are a primary contributor to the BH merger rate, at least at the high-end of the mass spectrum.

\begin{figure}
\includegraphics[width=\columnwidth]{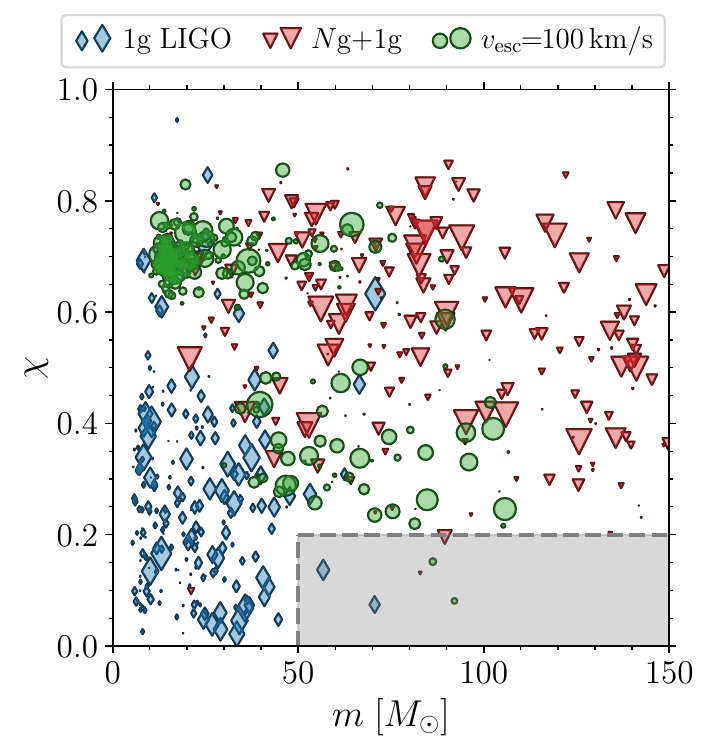}
\caption{Component BH masses $m$ and spins $\chi$ for GW events formed in hierarchical-merger models. Blue diamonds indicate a population of first-generation sources compatible with current LIGO data. From these, we construct higher-generation BHs considering both $N$g+1g assemblies (red triangles, Sec.~\ref{ng1grunaway}) and  hosts with fixed escape speed $v_{\rm esc}=100~{\rm km\,s^{-1}}$ (green circles, Sec.~\ref{vesc}). Hierarchical mergers cannot efficiently populate the gray region in the bottom right corner. The few outliers from the LIGO population are due to the fact that the fit by  \cite{2021ApJ...913L...7A} includes GW190521;  those coming from the hierarchical-merger models have very specific properties (Sec.~\ref{slowremnant}). To avoid cluttering, we only show the high-generation BHs involved in the hierarchical mergers, not their first-generation companions extracted from the LIGO distribution. The size of the markers is linearly proportional to the LIGO detectability (see the text for details).}
\label{oneghigh}
\end{figure}

\begin{figure*}
\includegraphics[width=\textwidth]{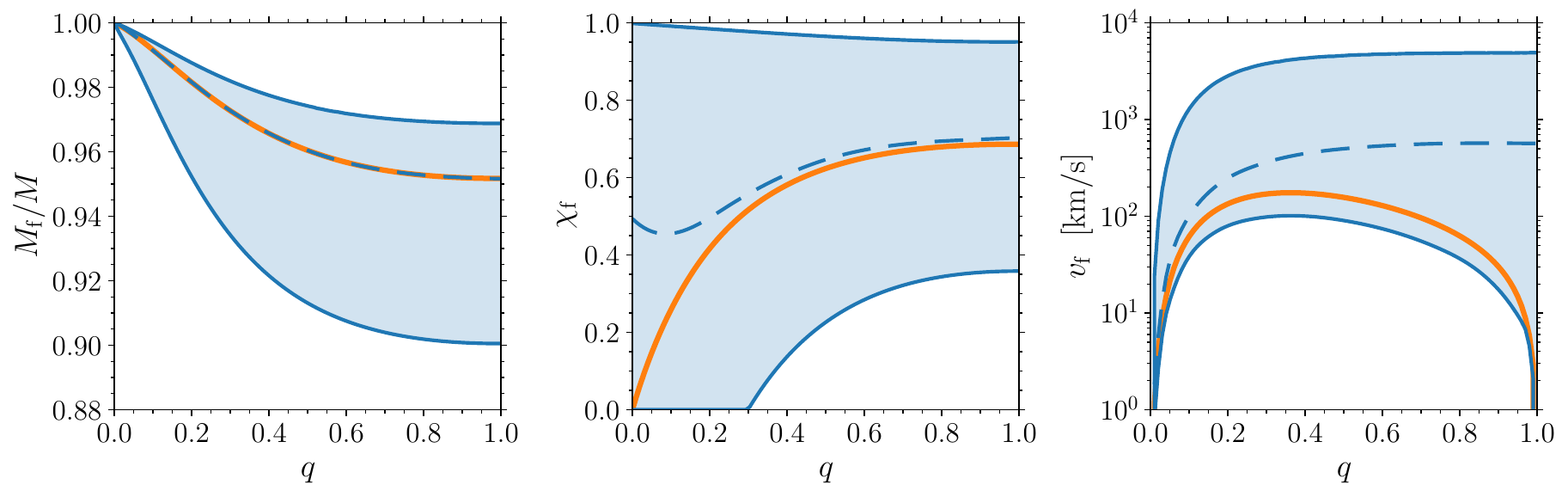}
\caption{Allowed values of final mass $M_{\rm f}$ (left panel, in units of the binary total mass $M$), final spin $\chi_{\rm f}$ (middle panel), and final recoil $v_{\rm f}$ (right panel) as a function of the binary mass ratio $q$.  The solid blue curves and the shaded areas show the largest and smallest remnant value if one extremizes over all 6 degrees of freedom of the two component spin vectors. The dashed blue curves show the median values for a spin distribution with uniform spin magnitudes $\chi_{1,2}\in[0,1]$ and isotropic spin directions. The orange thick curves mark the remnant properties resulting from the merger of two Schwarzschild BHs, i.e. $\chi_1=\chi_2=0$.}
\label{remnantfits}
\end{figure*}

The key idea behind this paper is presented in Fig.~\ref{oneghigh} where we show masses $m$ and spins $\chi$ of the BH-binary components formed via hierarchical mergers and detectable by LIGO. We start from a population of BHs as  inferred by \cite{2021ApJ...913L...7A} using current GW data (blue diamonds). We then construct two simplified, but complementary scenarios to illustrate the existence of the high-mass/low-spin exclusion region. First, we consider higher-generation BHs  grown from individual seeds that accrete companions sequentially, one after the other (red triangles; Sec.~\ref{ng1grunaway}). This is applicable to, e.g., runaway collisions in clusters as well as BHs formed via gas-assisted migration in AGN disks. We also consider the hierarchical assembly of BHs in astrophysical hosts with fixed escape speed, mimicking the dynamics of globular and nuclear star clusters (green circles; Sec.~\ref{vesc}). In both cases and  independently of  our modeling assumptions, the gray region with $m\gtrsim 50 M_\odot$ and $\chi\lesssim 0.2$ remains essentially empty. Very large spins $\chi\gtrsim 0.9$ are also unlikely.

This paper is organized as follows. In Sec.~\ref{slowremnant}, we investigate the properties of BH binaries that form slowly rotating remnants. We then present our  two simplified models: $N$g+1g sequential mergers (Sec.~\ref{ng1grunaway}) and  hosts with fixed escape speed (Sec.~\ref{vesc}, Appendix~\ref{app}).  Finally, in Sec.~\ref{concl} we draw our conclusions. We use geometric units where $c=G=1$.

\section{Slowly rotating merger remnants}
\label{slowremnant}

The exclusion region we wish to highlight is a direct consequence of the properties of BH mergers. These tend, on average, to produce remnants with spins that are close to a characteristic value: $\chi\ssim 0.7$ \citep{2008ApJ...684..822B,2017PhRvD..95l4046G,2017ApJ...840L..24F,2021CQGra..38d5012G}.
In this section, we summarize the conditions that binaries need to satisfy to instead produce remnants with  lower spins~(see~\citealt{2006PhRvD..74d1501C,2007PhRvD..76f4034B,2008ApJ...674L..29R,2008PhRvD..77b6004B, 2020A&A...640L..20B}).

Throughout this paper, we compute the post-merger properties using fits to numerical-relativity simulations for final mass $M_{\rm f}$~\citep{2012ApJ...758...63B},  final spin $\chi_{\rm f}$~\citep{2016ApJ...825L..19H}  and   recoil $v_{\rm f}$~\citep{2016PhRvD..93l4066G}. Although the fits by \cite{2019PhRvR...1c3015V,2019PhRvL.122a1101V} are more accurate for binaries with mass ratios $q\equiv m_2/m_1 \gtrsim 0.25$, we opt for those earlier expressions because they explicitly include information on the test-particle limit, which is a crucial piece of physics for this paper.

Figure~\ref{remnantfits} shows the properties of the post-merger BH as a function of the binary mass ratio $q$, extremizing over the component spin vectors $\boldsymbol{\chi}_1$, and $\boldsymbol{\chi}_2$. The radiated energy $M_{\rm f} -M$ ranges from 0 to a maximum of $\ssim 10\%$.  A nonspinning remnant ($\chi_{\rm f}=0$) is possible only if $q\lesssim 0.3$, while the less stringent condition 
$\chi_{\rm f}\lesssim 0.2$ requires $q\lesssim 0.5$. This is because, for comparable mass binaries, the orbital angular momentum at plunge $L\propto q/(1+q)^2$ is too large to be counterbalanced by the spins of the two merging BHs. For a set of BH binaries with spins distributed uniformly in magnitude and isotropically in direction, the median of $\chi_{\rm f}$ lies between $\ssim0.5$ and $\ssim 0.7$ for all values of $q$. The imparted kicks $v_{\rm f}$ range from 0, for highly symmetric configurations \citep{2008PhRvL.100o1101B}, to $O(1000)~{\rm km\,s^{-1}}$ (``superkicks'', see~\citealt{2007ApJ...659L...5C,2007PhRvL..98w1101G}).

\begin{figure}
\includegraphics[width=\columnwidth]{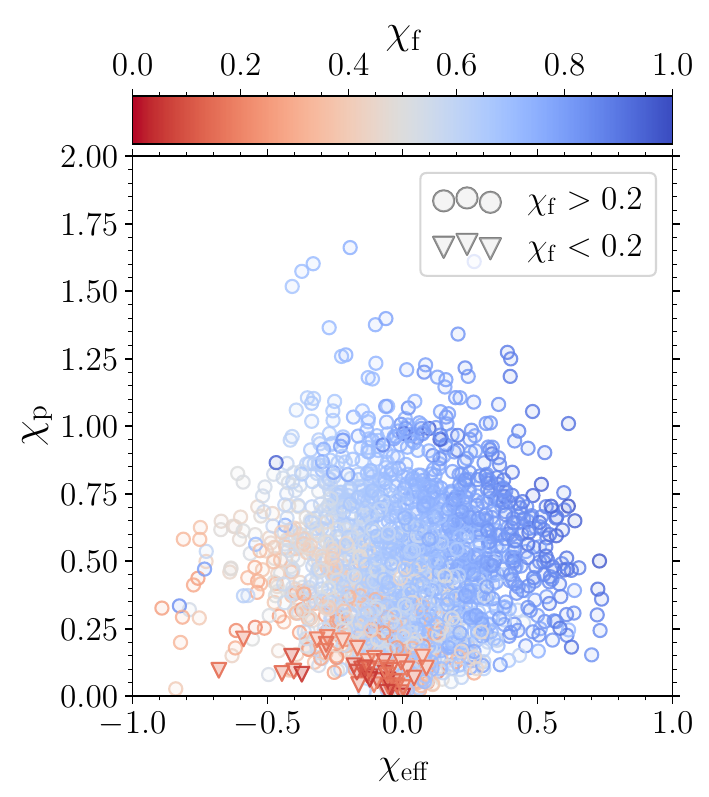}
\caption{Spin properties of BH binaries and their remnants. The $x$ and $y$ axes report the effective spins $\chi_{\rm eff}$ and $\chi_{\rm p}$, respectively. The spin of the post-merger remnant $\chi_{\rm f}$ is indicated on the color scale. This figure is produced assuming a population of BH binaries with uniform mass ratios $q\in [0,1]$, uniform spin magnitudes $\chi_{1,2}\in [0,1]$, and isotropic spin directions. Circles (triangles, plotted on top) mark binaries with $\chi_{\rm f}>0.2$ ($\chi_{\rm f}<0.2$). We use the precession-averaged  definition of $\chi_p\in[0,2]$ developed by \cite{2021PhRvD.103f4067G}. }
\label{wheresmallspin}
\end{figure}

Figure~\ref{wheresmallspin} contrasts the spins of the merging binary to that of the remnant. For this example, we consider a distribution of sources with uniform mass ratios $q\in[0,1]$, uniform spin magnitudes $\chi_{1,2}\in[0,1]$, and isotropic spin directions, and fix the orbital separation to $r=10 M$ (where $M$ is the total mass). For this population, the fraction of binaries with $\chi_{\rm f}\lesssim 0.2$ ($\chi_{\rm f}\lesssim 0.1$) is $\ssim 0.5\%$ ($\ssim 0.1\%$). For reference, the region $\chi_{\rm f}<0.2$ is excluded from the 90\% confidence interval of all current GW events \citep{2019PhRvX...9c1040A,2020arXiv201014527A}. The system with the lowest final spin is GW190814 ---probably a population outlier~\citep{2021ApJ...913L...7A}--- which has $\chi_{\rm f}\sim 0.28$. All other events present~$\chi_f\gtrsim 0.6$~\citep{2019PhRvX...9c1040A,2020arXiv201014527A}. Extrapolating from the entire population of LIGO sources, \cite{2021ApJ...914L..18D} estimated that the rate of remnants with $\chi_{\rm f }
\lesssim 0.2$ is at least three orders of magnitude lower compared to $\chi_{\rm f}\sim 0.7$.

In particular, Fig.~\ref{wheresmallspin}  translates the $\chi_{\rm f}$ constraints in terms of the effective spin parameters $\chi_{\rm eff}$ \citep{2008PhRvD..78d4021R} and $\chi_{\rm p}$~\citep{2015PhRvD..91b4043S}, which are often used to characterize GW events. In this paper, we use the precession-averaged expression of $\chi_{\rm p}\in[0,2]$ as derived by \cite{2021PhRvD.103f4067G}, which rectifies an inconsistency in the more commonly used definition.

Post-merger remnants with spin $\chi_{\rm f}\lesssim 0.2$ are produced almost exclusively by binaries with $\chi_{\rm eff}\lesssim 0$ and $\chi_{\rm p}\lesssim 0.25$. However, this is a necessary but not sufficient condition: the bottom left corner of the ($\chi_{\rm eff}$--$\chi_{\rm p}$) plane is also populated by a more numerous set of binaries with larger values of $\chi_{\rm f}$. For the specific population analyzed here, we find that only $\ssim$15\% ($\ssim4$\%) of the binaries with  $\chi_{\rm eff}< 0$ and $\chi_{\rm p}< 0.25$ presents $\chi_{\rm f}<0.2$ ($\chi_{\rm f}<0.1$). 

Although the precise fractions depend on the assumed population, our discussion  indicates that, overall,  slowly rotating BHs are a rare and fine-tuned outcome of BH mergers. 
We now investigate their production using two toy models.

\begin{figure*}[p]
\includegraphics[width=0.92\textwidth]{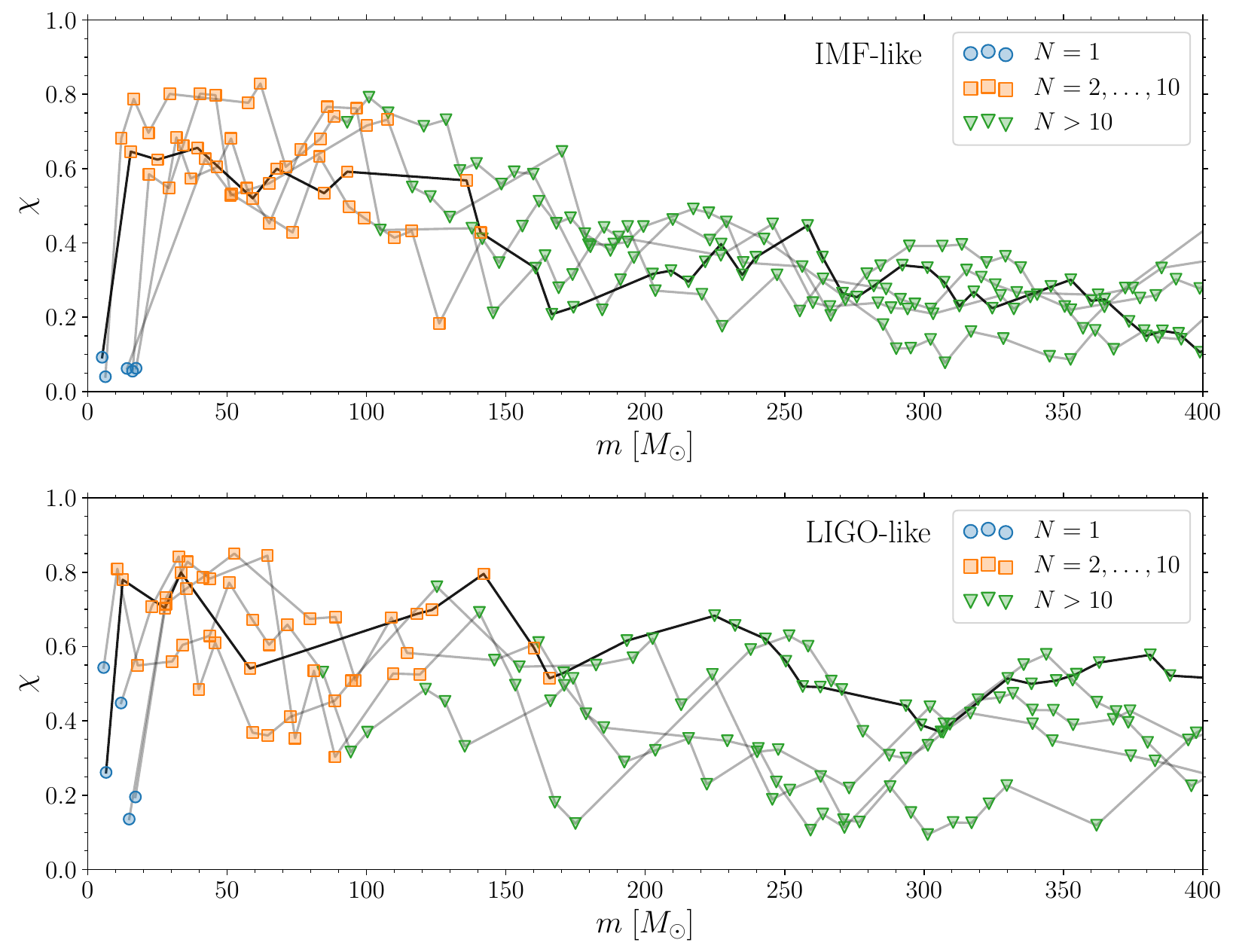}
\caption{Representative evolutionary tracks for the masses $m$ and spins $\chi$ of the $N$g BHs  in our runaway merger model. The top (bottom) panel uses the IMF-like (LIGO-like) 1g population. The initial seeds ($N=1$) are marked with blue circles. The first few merger generations ($2\leq N \leq 10$) are marked with orange squares. The rest of the evolution ($N>10$) is indicated with green triangles. To guide the eye, one particular track per panel is highlighted in black.}
\label{fewtrackssimple}
\bigskip\bigskip
\includegraphics[width=0.92\textwidth]{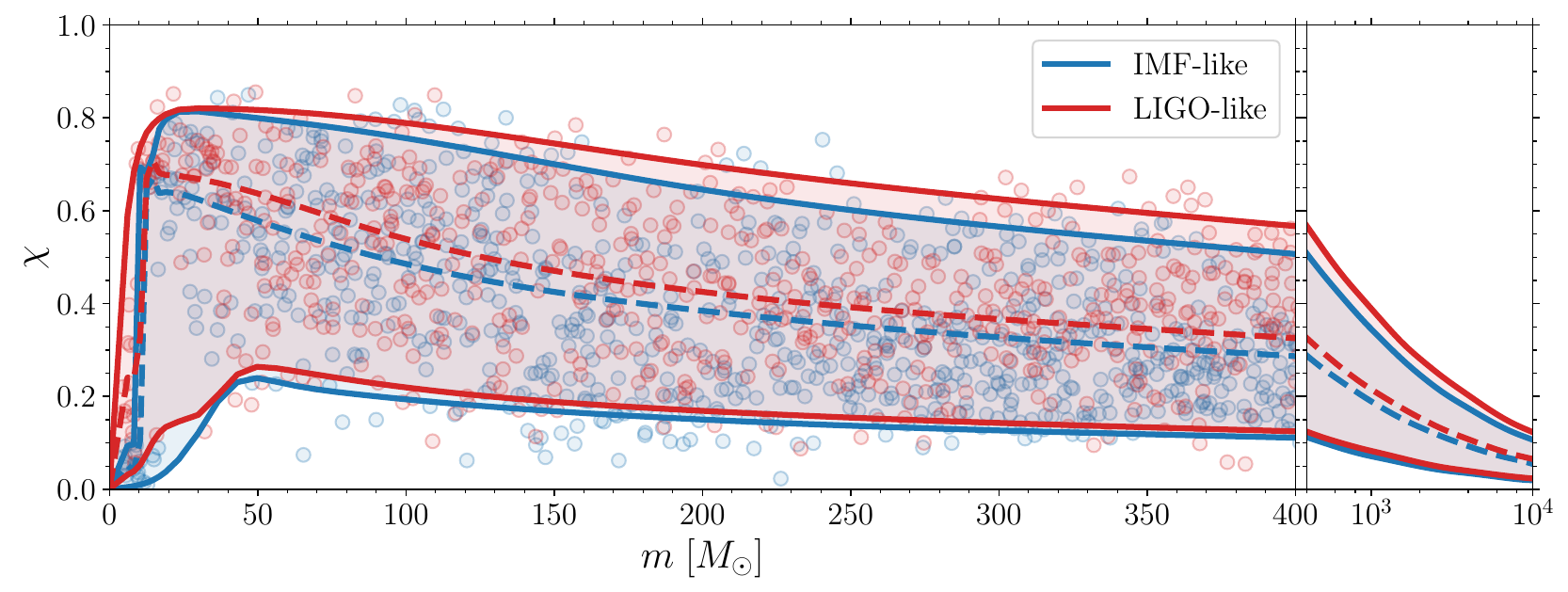}
\caption{Distribution of masses $m$ and spins $\chi$ of the $N$g BHs in our runaway collision model. Evolutions performed assuming the IMF-like (LIGO-like) population are indicated in blue (red). The shaded areas encompass $90\%$ of the spin values for each mass bin. Solid and dashed curves indicate the 5th, 50th, and 95th  percentiles of the distributions. Scatter points indicate a representative subset of BHs.}
\label{both90int}
\end{figure*}

\section{Ng+1g hierarchical assembly}
\label{ng1grunaway}

In this model, we consider a sequence of $N$g+1g encounters (hereafter ``g'' stands for generation). 
The basic idea is that, among the many BHs present in the environment, one object starts growing more than the others and ultimately accretes most or  all of its companions. This is a toy model to mimic runaway collisions in dense stellar clusters, which have long been invoked to describe the formation of intermediate-mass BHs
 while the first iterations are relevant to LIGO and the pair-instability mass gap \citep{1989ApJ...343..725Q,2002MNRAS.330..232C,2002ApJ...566L..17M,2019MNRAS.486.5008A,2021MNRAS.502.2682A,2021arXiv210305016M}. In AGN disks, the presence of migration traps \citep{2016ApJ...819L..17B} is also expected to drive the assembly of similar $N$g+1g merger chains~\citep{2019PhRvL.123r1101Y}

First, we select a seed BH from a prescribed population of 1g objects. We then construct a series of mergers of this seed with other BHs extracted from the same distribution. At each step, the generation $N$ of the highlighted BH increases and so does its mass. The evolution of the spin is less trivial.

We tested a variety of 1g BH distributions and found our results to be largely independent of the injected prescription. For concreteness, we present results  obtained under two sets of assumptions:
\begin{itemize}
\item {\it IMF-like}. This is our simplest scenario, where we extract BH masses from a power-law distribution $p(m)\propto m^\alpha$ with  $m\in[5M_\odot,50 M_\odot]$. The spectral index $\alpha=-2.3$ is modeled after  the stellar  initial mass function \citep{2001MNRAS.322..231K}. The sharp cutoff at $50M_\odot$ models the occurrence of pair-instability supernovae~\citep{2019ApJ...887...53F, 2020MNRAS.493.4333R}.  BH spins $\chi$ are uniformly distributed in $[0,0.1]$, which is inspired by scenarios where strong core-envelope interactions in massive stars  lead to slowly rotating remnants~(\citealt{2019ApJ...881L...1F,2020A&A...636A.104B}, see also \citealt{2016MNRAS.458.3075A}). This a conservative assumption that ensures that the critical region of interest ($\chi\lesssim 0.2$) is abundantly populated at least in the 1g distribution. The numerical-relativity fitting formulae are evaluated assuming isotropic spin directions.

\item {\it LIGO-like}. In this case, we make use of phenomenological fits to current LIGO/Virgo data. We extract BH binaries from the population posterior distribution derived by \cite{2021ApJ...913L...7A}, assuming their ``power-law$+$peak'' mass model and ``default'' spin model. Compared to the simpler IMF-like model, this 1g population presents a smooth mass cutoff between $m\sim 40 M_\odot$ and $
\sim 70 M_\odot$, and a distribution of component spins that extends up to $\chi\lesssim 0.5$ (see Fig.~
\ref{oneghigh}).  Sampling this population provides two sets of values $(m_i,\chi_i,\theta_i)$ for both the primary ($i=1$) and the secondary ($i=2$) BH component (where $\theta_i$ is the angle between the BH spin of BH $i$ and the orbital angular momentum of the binary).  At the first iteration, the seed of our $N$g+$1$g merger series is selected by randomly selecting either $i=1$ or $i=2$. We assume that the first merger ($N=1$) happens with the prescribed companion member ($i=2$ if we had previously selected $i=1$, and vice versa). This ensures that the resulting 1g+1g mergers are compatible with current LIGO/Virgo data. At each subsequent step ($N\geq2$), the mass and spin magnitude of the $N$g BH are inherited from the previous generation and its polar spin angle $\theta$ is assumed to be isotropic. The values of $m,\chi,$ and $\theta$ of the 1g companion are extracted from the same LIGO fits, once more randomly selecting $i=1,2$. The azimuthal spin angles are not modeled by \cite{2021ApJ...913L...7A} and are thus assumed to be isotropically distributed (this is equivalent to the Bayesian prior used in underlying the single-event analyses; \citealt{2019PhRvX...9c1040A,2020arXiv201014527A}).
\end{itemize}

Our results are presented in Figs.~\ref{fewtrackssimple} and \ref{both90int}. In particular, Fig.~\ref{fewtrackssimple} shows a small set of evolutionary tracks for illustrative purposes, while in Fig.~\ref{both90int} we average over many realizations. In both figures, we only show the $N$g BHs as they evolve because of mergers, not the 1g companions. %

Each evolution starts with a 1g seed, marked with blue circles in Fig.~\ref{fewtrackssimple}. The BH masses $m$ and spins $\chi$ evolve as additional 1g objects are accreted one after the other. The mass $m$ always increases with the generation $N$, which implies that Figs.~\ref{fewtrackssimple} and  \ref{both90int} can be read thinking that time flows, perhaps not uniformly, from left to right. Mergers with higher $N$ have a progressively lower mass ratio $q$.

The value of the spin $\chi$ jumps to $\ssim 0.7$ after just $\ssim 1$ merger generation. Any memory of the spin of the seed is lost very efficiently, which is a known property of BH mergers~\citep{2008ApJ...684..822B,2017PhRvD..95l4046G,2017ApJ...840L..24F,2021CQGra..38d5012G}. In $\lesssim 5$ generations, the mass of the BH surpasses the injected 1g cutoff of $\sim 50 M_\odot$ and enters the pair-instability gap. Crucially, while the mass of the BH is $50 M_\odot \lesssim m \lesssim 100 M_\odot$ (orange squares in Fig.~\ref{fewtrackssimple}), the spin remains relatively high $(\chi\gtrsim 0.4)$. Figure \ref{both90int} shows that the region of the parameter with $50 M_\odot \lesssim m \lesssim 100 M_\odot$ and $\chi\lesssim 0.2$ is excluded at the $\ssim 95\%$ level for both the IMF-like and the LIGO-like models.

After $\ssim 10$ generations (green triangles in Fig.~\ref{fewtrackssimple}), the BHs enter the proper intermediate-mass regime $m\gtrsim 100M_\odot$ and the spin starts to decrease. In particular, we find that $\chi\lesssim 0.8$ ($\chi\lesssim 0.4$) for $m\gtrsim 100 M_\odot$ ($m\gtrsim 1000M_\odot$) at the $\ssim 95\%$ level (Fig.~\ref{both90int}).

 The large-$m$  limit of our $N$g+1g merger chains corresponds to events with progressively lower mass ratio $q$. In this test-particle regime, %
 incoming 1g BHs that are corotating (counter-rotating)  with the $N$g BH tend to increase (decrease) its spin. Although these two outcomes are equally likely for isotropic encounters, their effects do not cancel out because counter-rotating orbits are more efficient at extracting angular momentum than corotating orbits are at depositing it \citep{2003ApJ...585L.101H}.  This is the  same  behavior found in  the so-called ``chaotic accretion''  model of supermassive-BH growth \citep{2006MNRAS.373L..90K}. 

\begin{figure*}
\includegraphics[width=\textwidth]{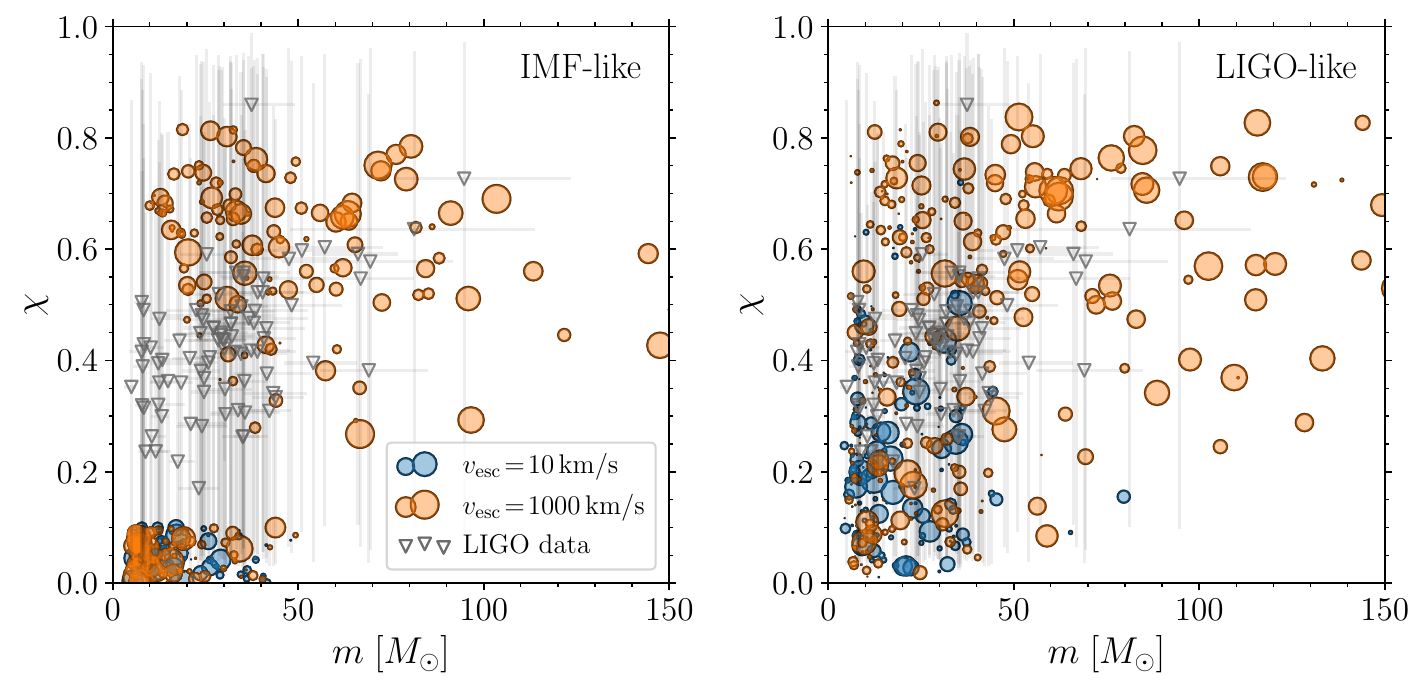}
\caption{Distribution of component masses $m$ and component spin magnitudes $\chi$ in our hierarchical-merger models with fixed escape speed. We show results for $v_{\rm esc}=10{\rm km\,s^{-1}}$ (blue) and $v_{\rm esc}=1000{\rm km\,s^{-1}}$ (orange) assuming either the IMF-like (left panel) or the LIGO-like (right panel) initial population. The size of the circles is linearly proportional to the LIGO detection probability  $p_{\rm det}$. The gray triangles and the associated error bars indicate medians and 90\% marginalized confidence intervals of 44 BH-binary events from LIGO/Virgo (thus resulting in $44\times 2=88$ component BHs).}
\label{individualmchi}
\end{figure*}
For a short proof, let us Taylor-expand the expression of  the post-merger BH spin  by \cite{2016ApJ...825L..19H} to first order $q$: 
\begin{align}
\chi_f &= \chi + q \left[\frac{f(\chi\cos\theta)}{\chi} -2\chi\right] +\mathcal{O}(q^2)\,.
\end{align}
Here $\chi$ and $\theta$ indicate the spin of the $N$g primary, and the function
\begin{align}
f(\chi') = \chi' \Big| L_{\rm ISCO}(\chi') - 2 \chi' [E_{\rm ISCO}(\chi')-1] \Big|
\end{align}
depends on the specific energy   $E_{\rm ISCO}(\chi')$  and angular momentum   $L_{\rm ISCO}(\chi')$ of a particle located at the innermost stable circular orbit (ISCO)  of a BH with spin $\chi'$ \citep{1973blho.conf..215B}.
Averaging over many merger events with isotropic spin directions [i.e. $\cos\theta\sim \mathcal{U}(-1,1)$] yields
\begin{align}
\langle\chi_f \rangle &= \chi + q \left[\frac{1}{2 \chi^2} \int_{-\chi}^{\chi}\!\! f(\chi')d\chi' - 2\chi\right] +\mathcal{O}(q^2)\,.
\end{align}
It is straightforward to show that the term in square brackets is $\leq 0$ for all values $0\leq \chi \leq 1$, which implies $\langle \chi_{\rm f}\rangle < \chi$ to first order in $q$. Because the mass of the $N$g BH increases as the merger chain proceeds, this statement is equivalent to $\langle d \chi/d m \rangle < 0$, which explains the asymptotic behavior observed in Fig.~\ref{both90int}.

To summarize, our simple $N$g+1g hierarchical model predicts a dearth of slowly rotating ($\chi\lesssim 0.2$) BHs in the pair-instability mass gap ($50 M_\odot \lesssim m\lesssim 100M_\odot$). Intermediate-mass BHs ($m\gtrsim 500 M_\odot$) with large spins ($\chi\gtrsim 0.6$) are also  rare. These statements are largely independent of the properties of the injected population.

\section{Hosts with fixed escape speed}
\label{vesc}

A trivial condition for BHs to merge repeatedly is that  remnants are retained in the astrophysical host following merger recoils. 
For our second model, the strategy is thus to simulate the assembly of hierarchical BH merger in dense stellar environments as a function of their escape speed. %
 In this toy experiment, a ``cluster'' is simply a collection of BHs characterized by a given escape speed~$v_{\rm esc}$.

Our algorithm proceeds as follows (see~\citealt{2019PhRvD.100d1301G,2020PhRvL.125j1103G} for previous applications). Given an initial population of BHs, we randomly select two objects and compute their remnant properties $M_{\rm f}, \chi_f$, and $v_{\rm f}$. If $v_{\rm f}>v_{\rm esc}$, the remnant is removed from the cluster. If instead $v_{\rm f}<v_{\rm esc}$, we place it back in the sample together with the other 1g objects. We then proceed by  randomly selecting two BHs from those that are left and iterate until $< 2$ objects remain.

The injected populations are the same as the ones we already introduced in Sec.~\ref{ng1grunaway}: 

\begin{itemize}
\item The IMF-like case is a straightforward application of the algorithm above. We start from $N$ individual BHs and, at each step,  select two objects with a flat pairing probability between any of the possible couples. This choice was dubbed ``random pairing''  by \cite{2019PhRvD.100d1301G} and is tentatively supported by current GW observations, which indicate that merging BHs do not prefer companions with a given mass or spin (see the best-fit value of the parameter $\beta_q$ found by \citealt{2021ApJ...913L...7A}, which is compatible with $0$, as well as inference performed with their ``multi spin'' model, which is inconclusive). All spin directions are distributed isotropically.

\item Sampling the LIGO-like model does not provide individual objects, but paired binaries. In this case, our initial collection is made of $N/2$ population draws. At the first iteration, we pick one of these pairs and consider its merger. This ensures that the final population of GW events presents a subsample (one event per cluster) with masses and spins that are compatible with current data. From the second merger onwards, we unpair all BHs from their companions and proceed with a randomly uniform selection as above. Each 1g BH carries the values of $m,\chi$, and $\theta$ extracted from the LIGO fit. Whenever a post-merger remnant reenters the cluster following a kick, its spin direction is resampled from an isotropic distribution.  
\end{itemize}

\cite{2019PhRvD.100d1301G} showed that the properties of the resulting mass-gap binaries converge as long as the initial number of BHs is $N\gtrsim 1000$; we thus set $N=5000$. For any given value of $v_{\rm esc}$, we run multiple cluster realizations to  obtain statistically solid samples.
For each recorded GW event, we estimate its detection probability $p_{\rm det}\in [0,1]$ using the semianalytic approach by \cite{1993PhRvD..47.2198F} and \cite{1996PhRvD..53.2878F}. For simplicity, selection effects are computed neglecting spins (a very marginal effect; see \citealt{2018PhRvD..98h3007N}), assuming a redshift distribution $z\in[0,1]$ that is uniform in comoving volume and source-frame time, filtering out systems with total mass $M>300 M_\odot$, considering the noise-curve of LIGO at design sensitivity \citep{2020LRR....23....3A}, the waveform model by \cite{2016PhRvD..93d4007K}, and a single-detector signal-to-noise ratio (SNR) threshold $\rho=8$ \citep{2016ApJS..227...14A}.  The same assumptions were used to weight the BHs illustrated in Fig.~\ref{oneghigh} for both models.

Figure \ref{individualmchi} shows the resulting distributions of component BHs for $v_{\rm esc}=10{\rm km\,s^{-1}}$ and $v_{\rm esc}=1000{\rm km\,s^{-1}}$. With typical BH merger kicks of $O(100){\rm km\,s^{-1}}$ (e.g. \citealt{2007ApJ...662L..63S,2012PhRvD..85h4015L,2019PhRvD.100d1301G}), these two models bracket the overall phenomenology. The intermediate case with $v_{\rm esc}=100~{\rm km\,s^{-1}}$ is presented in Fig.~\ref{oneghigh}. Additional model variations are explored in Appendix~\ref{app}. For comparison, the typical escape speed of a globular cluster is $\ssim 30~{\rm km\,s^{-1}}~{\rm km\,s^{-1}}$ (\citealt{2002ApJ...568L..23G,2004ApJ...607L...9M}, reaching up to  $\ssim 100~{\rm km\,s^{-1}}$ in some cases), those of nuclear star clusters are up to an order of magnitude larger \citep{2009ApJ...692..917M, 2016ApJ...831..187A}, and large elliptical galaxies reach $v_{\rm esc}\gtrsim 1000~{\rm km\,s^{-1}}$ \citep{2004ApJ...607L...9M, 2015MNRAS.446...38G}.

In the low escape-speed case (blue circles in Fig.~\ref{individualmchi}), our model returns a single population that is almost entirely made of 1g+1g BHs. If instead $v_{\rm esc}$ is larger than the typical kicks (orange circle), there is an additional high-g population of BHs. These hierarchically formed BHs efficiently populate the pair-instability mass gap ($m\gtrsim 50 M_\odot$) but, crucially, they do so with large spins ($\chi\gtrsim 0.2$). The bottom right corners of both panels in  Fig.~\ref{individualmchi} remain essentially empty, independently of $v_{\rm esc}$.    More specifically, spins are marginally higher for $v_{\rm esc}=1000~{\rm km\,s^{-1}}$ (Fig.~\ref{individualmchi}) compared to $v_{\rm esc}=100~{\rm km\,s^{-1}}$ (Fig.~\ref{oneghigh}) because final spins and final kicks tend to be positively correlated (Sec. \ref{slowremnant}, see also \citealt{2021ApJ...914L..18D} for an application with LIGO data).

In the LIGO-like case (see Figures~\ref{oneghigh} and \ref{individualmchi}), there are a few outliers with $50 M_\odot \lesssim m \lesssim 70 M_\odot$ and $\chi\lesssim 0.2$. One must keep in mind that, in this model, we are effectively assuming that the entire LIGO/Virgo dataset is made of 1g BHs. Because of GW190521, the injected 1g mass spectrum has some support in the  $m\gtrsim 50\ M_\odot$ region. A more refined estimate of the lower limit of the exclusion region should consider removing the most massive events from the catalog, subject to the consideration that they could already be of higher generation.

\begin{figure*}[p]
\includegraphics[width=\textwidth]{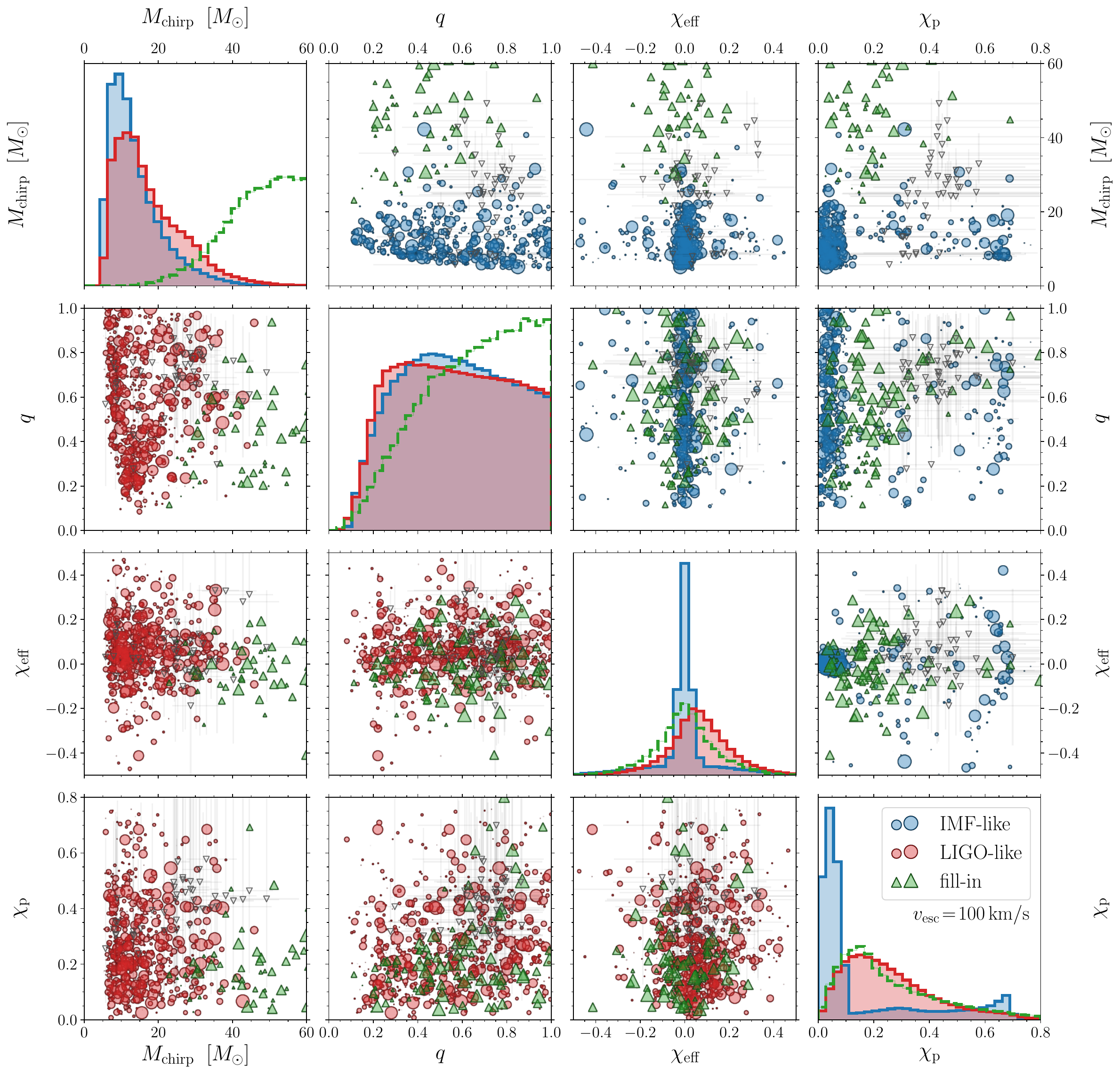}
\caption{Joint distribution of chirp mass $ M_{\rm chirp}$, mass ratio $q$, effective spin $\chi_{\rm eff}$, and precession parameter $\chi_{\rm p}$ assuming clusters with $v_{\rm esc}=100~{\rm km\,s^{-1}}$. Results obtained with the IMF-like and LIGO-like distributions are indicated in blue and red, respectively. The size of the markers  and the entries in the marginalized histograms are linearly weighted by the LIGO detectability $p_{\rm det}$. The green distributions (triangular markers, dashed histograms) indicate a population that has been artificially designed to populate the pair-instability mass gap with slowly rotating BHs.
}
\label{fourcorner}
\end{figure*}

Because of large measurement errors, such an exclusion region is \emph{not} empty observationally.  Figure \ref{individualmchi}  shows the measured values of $m$ and $\chi$ for some the current LIGO/Virgo events, together with their uncertainties. In particular, we consider\footnote{For the events detected during the  first and second  observing runs, we use posterior samples released by \cite{2020MNRAS.499.3295R}. For the first half of the third observing run, we use the samples by \cite{2020arXiv201014527A} labeled as ``publication'' and ``comoving"; see details therein.} the 44 BH-binary events
reported by \cite{2019PhRvX...9c1040A,2020arXiv201014527A} with false-alarm rate $<1$yr and used in the analysis by \cite{2021ApJ...913L...7A}.
A few of the observed component BHs are compatible with  masses $m\gtrsim 50 M_\odot$ and spins $\chi\lesssim 0.2$. 

GW detectors observe binaries, not individual BHs. Disentangling the two components from the data is, at present, subject to large uncertainties, especially for the spins. Current measurement errors are as large as $\Delta\chi\sim 1$ and, unfortunately, are not expected to improve significantly with the current facilities \citep{2016PhRvD..93h4042P,2017PhRvD..95f4053V}. The accuracy on the spins decreases with the mass because fewer and fewer inspiral cycles take place in band.  This implies that, from a data-analysis perspective, the high-mass/low-spin region we are interested is one of the most challenging corners of the parameter space. Third-generation detectors like the Einstein Telescope~\citep{2010CQGra..27s4002P} and Cosmic Explorer~\citep{2019BAAS...51g..35R}  will be a game changer. With SNR larger by about a factor of 100, those future machines will reach accuracies  $\Delta\chi\sim 10^{-2}$  (see \citealt{2017PhRvD..95f4052V}, but note they only analyze GW frequencies $>10$ Hz, thus penalizing spin effects that accumulate early in the inspiral). This will allow us to confidently place individual mass-gap BHs either inside or outside the $\chi\lesssim 0.2$ region.

The parameters that are best measured in GW observations include the chirp mass $M_{\rm chirp}$, the mass ratio $q$, the effective spin $\chi_{\rm eff}$, and the precession parameter $\chi_{\rm p}$. Figure \ref{fourcorner} shows the distributions of these four quantities for our two models with $v_{\rm esc}=100~{\rm km\,s^{-1}}$, as well as the current LIGO/Virgo events.  The values of $\chi_{\rm p}$ are computed at the reference GW frequency $f=20$ Hz, with the exception of LIGO's GW190521 for which $f=11$~Hz. 

In terms of these binary  estimators, the exclusion region is still present but becomes harder to discern. To better exemplify this point, in Fig.~\ref{fourcorner} we include an additional population (``fill-in'') where all binaries have, by construction, a slowly rotating primary BH in the pair-instability mass gap. In particular, we distribute $m_1$ uniformly in $[50 M_\odot,150 M_\odot]$, $\chi_1$ uniformly in $[0,0.2]$, $m_2$ uniformly in $[5 M_\odot,m_1]$, $\chi_2$ uniformly in $[0,1]$, and $z$ uniformly in comoving volume and source-frame time in $[0,1]$, and  assume isotropic spin directions. Compared to both the IMF-like and LIGO-like case with hierarchical mergers, the fill-in distribution preferentially populates the region with $M_{\rm chirp}\gtrsim 40 M_\odot$ and $\chi_{\rm p}\lesssim 0.3$. 

Even if the accuracy of the properties of the individual BHs is, at present at least, too poor to exploit the exclusion region, the results presented in Fig.~\ref{fourcorner} leave open the possibility of ruling out the occurrence of hierarchical mergers statistically at a population level. Early steps in this direction are already being pursed, see, e.g., the ``multi spin'' model by~\cite{2021ApJ...913L...7A} and the	``pollutant'' population by~\cite{2021arXiv210402685B}. Recent models by \cite{2021arXiv210409510T}  also indicate that hierarchical mergers are unlikely to form GW events with large chirp mass and low effective spins. Even in this admittedly less clean case, our results highlight the importance of combining mass and spin information ($M_{\rm chirp}$ and $\chi_{\rm p}$ in this case).

\section{Conclusions}
\label{concl}

The LIGO/Virgo detection of BHs with masses larger than $\ssim 50 M_\odot$ raises questions about our current understanding of compact-binary formation. Unstable pair production makes the formation of such massive BHs challenging to explain in conventional stellar evolution. Hierarchical BH mergers are among the proposed mechanisms to evade this limit~\citep{2017PhRvD..95l4046G,2017ApJ...840L..24F}, but other strategies of astrophysical nature have also been proposed. These include, for instance, accretion events in close binaries and stellar mergers (e.g.~\citealt{2019MNRAS.487.2947D,2019ApJ...882...36M}), as well as lowered nuclear-reaction rates and stellar rotation (e.g.~\citealt{2020A&A...640L..18M,2021MNRAS.501.4514C,2021ApJ...912L..31W}). 

We have pointed out that the BH parameter space presents a region that is exclusive to those additional astrophysical pathways. In particular, BHs with masses $\gtrsim 50 M_\odot$ and spins $\chi\lesssim 0.2$ are hard to explain with either conventional stellar collapse (because of the pair-instability limit) or hierarchical mergers (because they almost exclusively produce rapidly rotating BHs). We have exemplified this statement using two  complementary toy models (Secs.~\ref{ng1grunaway} and \ref{vesc}) and expect it to hold for generic formation pathways that  include hierarchical mergers.

A putative future detection in this high-mass/low-spin region of the parameter space will need to be explained with other astrophysical arguments: the ``easy way out'' of hierarchical merger will not be viable. The only escape route for hierarchical mergers to remain relevant  would require a considerable fine-tuning of the properties of the previous generation(s) of mergers~(Sec.~\ref{slowremnant}). In particular, the progenitor BH  binary will need to have had both sufficiently extreme mass ratio $q\lesssim 0.5$ and sufficiently small effective spins $\chi_{\rm eff}\lesssim 0,\chi_{\rm p}\lesssim 0.25$. These properties will then need to be statistically compatible with the other, far more numerous events in the GW catalog. If that turns out not to be the case, even this last possibility can be excluded and hierarchical mergers definitely  ruled out. For instance, in a scenario where all 1g BHs are born nonspinning \citep{2019ApJ...881L...1F}, explaining a massive outlier with spin $
\lesssim 0.2$ via hierarchical mergers would require that a substantial fraction of the population has $q
\lesssim 0.1$, which is extremely unlikely (see \citealt{2020ApJ...901L..39O} and references therein).

There is, however, an important caveat. The exclusion region is best formulated in terms of the mass and spins of the component BHs. These properties are, at least with current facilities,  challenging to measure~(\citealt{2016PhRvD..93h4042P,2017PhRvD..95f4053V}, but see \citealt{2021PhRvL.126q1103B} for a different take). Our argument can be recast in terms of the binary estimators $M_{\rm chirp}$ and $\chi_{\rm p}$, but assessing its relevance will probably require a more complex population analysis, rather than  single smoking-gun events. 

Our study further increases the science case of future third-generation GW interferometers. They will deliver measurements of the individual masses and spins at the $1\%$ level,
fully realizing the potential of the high-mass/low-spin exclusion region we have highlighted.

\acknowledgements
We thank Maya Fishbach for providing a script to sample the LIGO/Virgo population fits. We thank Ulrich Sperhake, Paolo Pani, Daria Gangardt, Matthew Mould, Vijay Varma, and Michela Mapelli for discussions.
We made use of public data products released by~\cite{2020MNRAS.499.3295R} and \cite{2020arXiv201014527A,2021ApJ...913L...7A}. The detection probability $p_{\rm det}$ was evaluated using the public code of \cite{gwdet}.
D.G. and N.G. are supported by European Union's H2020 ERC Starting grant No. 945155--GWmining, Leverhulme Trust grant No. RPG-2019-350, and Royal Society grant No. RGS-R2-202004. A.V. acknowledges the support of the Royal Society and Wolfson Foundation, and the UK Science and Technology Facilities Council through grant No. ST/N021702/1. Computational work was performed on the University of Birmingham BlueBEAR cluster, the Athena cluster at HPC Midlands+ funded by EPSRC grant No. EP/P020232/1 and the Maryland Advanced Research Computing Center (MARCC).

\appendix
\section{Additional model variations} \label{app}

\begin{figure*}[p]
\includegraphics[width=\textwidth]{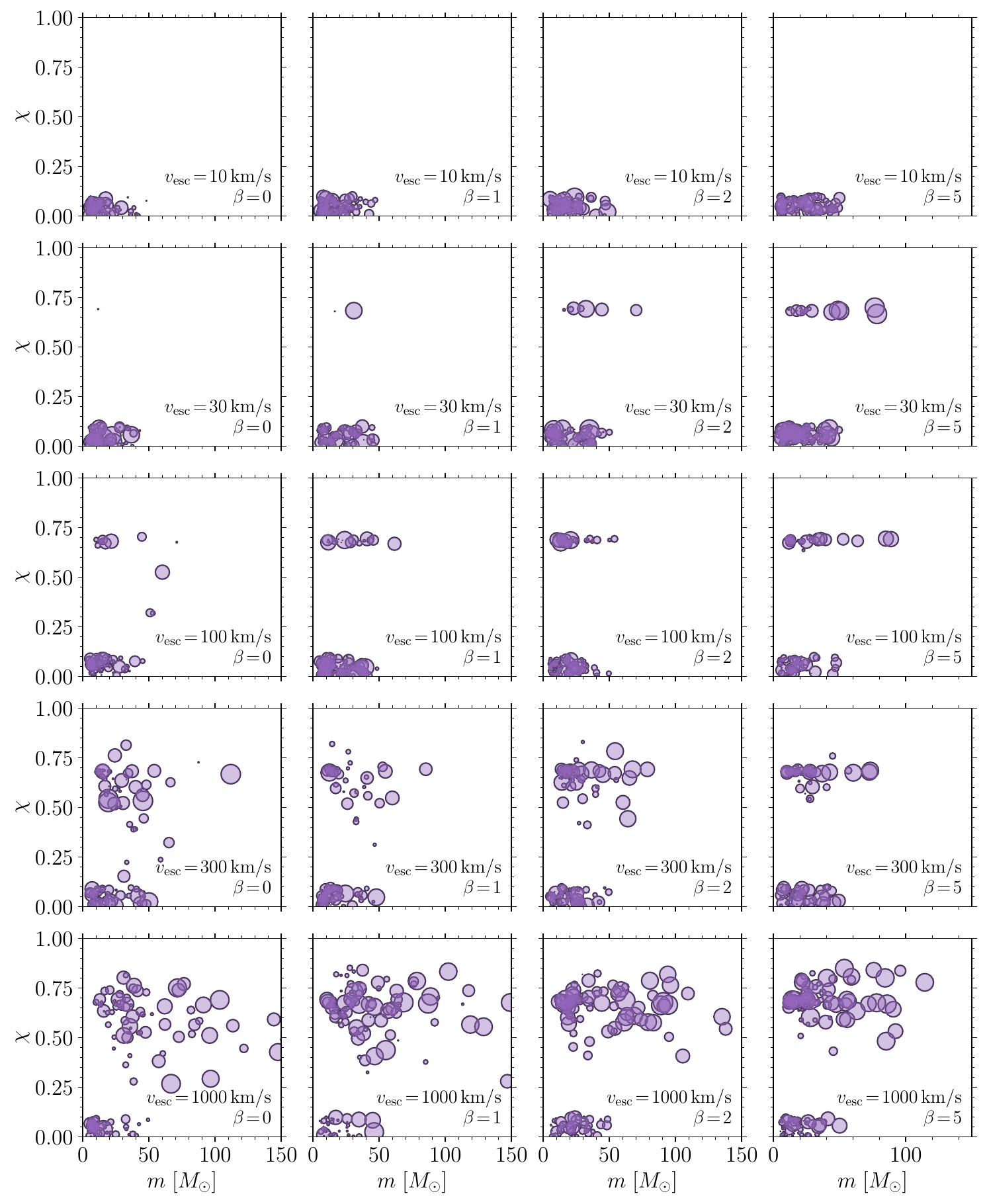}
\caption{Additional model variations expanding on the IMF-like model in hosts with fixed escape speed. We vary the escape speed of the environment $v_{\rm esc}$ (top to bottom) and the BH pairing spectral index $\beta$ (left to right). Markers indicate component masses $m$ and component spins $\chi$, and are resized linearly according to the LIGO detection probability $p_{\rm det}$.}
\label{changebeta}
\end{figure*}

Here we present additional variations expanding on the models  introduced in Sec.~\ref{vesc}. In particular, we consider the IMF-like case and investigate the impact of the BH pairing probability (see also \citealt{2019PhRvD.100d1301G}). At each step in the algorithm presented in Sec.~\ref{vesc}, we select two BHs from the cluster with a probability $p(q)\propto q^\beta$ where $\beta$ is a free parameter. The fiducial model presented in the main body of the paper corresponds to $\beta=0$ and describes an environment where merging BHs do not prefer companions with specific properties. Cases with $\beta>0$ instead describe scenarios where BHs pair preferentially with other objects of similar masses as predicted in, e.g., mass-segregated clusters \citep{2013LRR....16....4B}.

Our results are presented in Fig.~\ref{changebeta} where we consider $v_{\rm esc}=$ 10 ${\rm km\,s^{-1}}$, 30 ${\rm km\,s^{-1}}$, 100 ${\rm km\,s^{-1}}$, 300 ${\rm km\,s^{-1}}$, 1000 ${\rm km\,s^{-1}}$ and $\beta=$ 0, 1, 2, 5. %
The escape velocity mainly determines the rate of hierarchical mergers. As $v_{\rm esc}$ increases, the subpopulation of objects with $\chi\sim 0.7$ becomes more pronounced. The pairing spectral index $\beta$ has a subdominant effect and mainly determines the dispersion of the spin distribution in the hierarchical-merger subpopulation. In particular, BHs from repeated mergers formed in hosts  with large (small) values of $\beta$ have spins that are more (less) tightly peaked around $\chi\sim 0.7$. Larger values of $\beta$ imply that the resulting GW events have mass ratios closer to unity. Those BH binaries are more similar to each other compared to their $\beta=0$ counterparts and thus present a spin distribution with lower dispersion.

\bibliographystyle{yahapj}

\bibliography{repeatedsmallspins}

\end{document}